\documentclass[aps,twocolumn,prl,showpacs,nofootinbib]{revtex4}
\usepackage{amsmath}
\usepackage{graphicx}
\usepackage{dcolumn}
\usepackage{bm}
\usepackage{amssymb}

\begin{document}
\title{Ashkin-Teller formalism for elastic response of DNA
molecule to external force and torque}
\author{Zhe Chang}
\affiliation{Institute of High Energy Physics,Chinese Academy of
Sciences, P.O.Box918(4), 100049 Beijing, China }
\email{changz@mail.ihep.ac.cn}
\author{Ping Wang and Ying-Hong Zheng}
\affiliation{Institute of High Energy Physics, Chinese Academy of
Sciences, P.O.Box 918(4), 100049 Beijing, China,\\ Graduate
School, Chinese Academy of Sciences, 100049 Beijing, China}
\email{pwang@mail.ihep.ac.cn} \email{yh_zheng@mail.ihep.ac.cn}
\begin{abstract}
We propose an Ashkin-Teller like model for elastic response of DNA
molecule to external force and torque. The base-stacking interaction
is described in a simple and uniform way. We obtain the phase
diagram of dsDNA, and in particular, the transition from B form to
the S state induced by stretching and twisting. The elastic response
of the ssDNA is presented also in a unified formalism. The close
relation of dsDNA molecule structure  with elastic response is shown
clearly. The calculated folding angle of the dsDNA molecule is
$59.2^o$.
\end{abstract}
 \maketitle
\section{I. Introduction}
The DNA molecule is the basic genetic material. The ability of DNA
to pack and fold into chromosomes or to serve as a template during
transcription and replication depends on the particular elastic
properties of the molecule as modified by local interactions.
Attribute to the rapid development of the technique of directly
manipulating single molecule\cite{sm1,sm2,cl,strick,rief,cla},
measurement of elastic properties of single DNA molecule under
stretching or twisting becomes possible recently. An interesting
new phenomena revealed by experiments is that there exists a new
state, called S-state which exhibits exotic elastic properties
distinguished from the B state. The dsDNA molecule can be extended
to almost as twice as its original length before transits to the
new state. There is a sharp transition between the B-state and the
S-state, which shows an evidence of high level of cooperativity of
molecules in the dsDNA. Authors have argued that there must be
unknowing dsDNA molecular configuration corresponding to the
structural transition.

References\cite{sm2,st,le,mar1,mar2,kam,bou,mar3,rou1,rou2,tam,zhou1,zhou2,nel,aa,co}
provided valuable insights. However, discrepancy (for example,
untwisted or helical) still exists. Several models were proposed to
fit data of the dsDNA molecule response to external force. The
freely jointed chain (FJC) model treats dsDNA simply as a polymer
joined freely by rigid units and neglect the base-stacking
interaction between two adjoint base pairs. It fits the data at low
force limit well for DNA molecule. However, the FJC model can not be
used  to describe the behavior of dsDNA molecule  which subjected to
stretching and twisting simultaneously. It is well-known that
twisting also plays an important role during the structural
transition process of dsDNA molecule. From a biological point of
view, torsional stress is indeed popular in the living cell and may
strongly influence dsDNA functioning. The phenomenological
model\cite{st}, which takes stretching and twisting both into
account, describes every state with five independent parameters and
trends no phase transition occurred in the course of stretching and
twisting.

In this paper, we  present an Ashkin-Teller like model  to describe
the elastic response of dsDNA molecule in a unified framework
(various states with the same parameter) and try to find the
structural transition mechanism  of dsDNA molecule.

\section{  II. The Ashkin-Teller formalism}
The FJC model describes dsDNA molecule simply as joined rigid units.
While stretching one end of the molecule along the direction of the
strand and fixing another end, one can deduce a force-extension
relation\cite{nel} as

\begin{equation}
\displaystyle
\langle\frac{z}{L_{tot}}\rangle=\coth\left(\frac{fb}{k_{B}T}\right)-\frac{k_T}{fb},
\end{equation}
where $b$ is effective unit persistent length and $z$ is
displacement of dsDNA molecule along the strand under external
force $f$.
\begin{figure}[h]
 \begin{center}
 \includegraphics[scale=0.8]{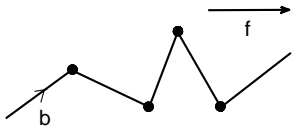}
 \caption[]{FJC model: freely jointed  rigid units.     }
 \end{center}
 \end{figure}
It is well-known that dsDNA is a double-stranded linear biopolymer,
each strand is a covalently linked chain of nucleotides. Nucleotides
come in four varieties and are composed of three distinct parts: a
sugar called deoxyribose, a phosphate group, and one of four units,
adenine, guanine, thymine, and cytosine called bases. Each of the
four bases attached to nucleotides is hydrophobic and capable of
hydrogen-bonding with another complementary base. Each arm of the
dsDNA is formed, in turn, by a covalent bond between the phosphate
group of one nucleotide and the hydroxyl group of its neighbor
nucleotide's sugar. The bond is often referred to as the
phosphodiester bond and the two arms collectively build up the DNA
backbone. This classical B-DNA structure is the basic dsDNA
conformation found inside living cells. Now, it is generally
accepted that contribution from base-stacking is the main reason
help to form steady two strand helix configuration, and it also
determines the dsDNA's local transformation. Unfortunately, the
precise picture about base-stacking is still unknown.

The FJC model treats the dsDNA molecule composed of two interacting
strands simply as the freely joined rigid units and neglects the
interaction between two adjoint bases (base-stacking). It is
suspected reasonably that it fails to describe high force extension
and can not take the torque effect into account. Here, we suggest an
Ashkin-Teller like model (based on the Ashkin-Teller
model\cite{at,ka} in statistical mechanics) to describe in a uniform
way the elastic response of DNA molecule to both external force and
torque.

The Hamiltonian  of the Ashkin-Teller like model  with external
force ($f$) and torque ($\sigma$) is of the form

\begin{multline}
\frac{H}{k_B T}=-\sum_i[s_i s_{i+1}+\tau_i \tau_{i+1}+K(f,\sigma)
s_i s_{i+1}\tau_i\tau_{i+1}\\+(\alpha f+\gamma)(s_i+\tau_i)]
,\label{H}
\end{multline}
where $s$ and $\tau$ are 1-dimension unit orientation vectors (takes
values of $\pm1$). $K(f,\sigma)$ describes the strength of
base-stacking interaction between two adjoint base pairs. In the
case of  $K(f,\sigma)=0$,  the model reduces to two decoupled
chains.
 $\alpha$ and $\gamma$ are constant parameters.
\begin{figure}[h]
 \begin{center}
 \includegraphics[scale=0.8]{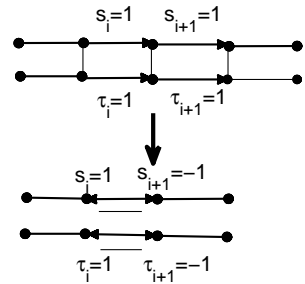}
 \caption[]{The  Ashkin-Teller like model consists of two coupled chains.
 A possible transform of configurations corresponds to elastic response to
 external interaction.  }
 \end{center}
 \end{figure}
It is obvious that the strength of base-stacking interaction depends
on structure of DNA molecule as well as external conditions in which
the DNA is studied. In general case of the DNA molecule under
stretching and twisting, we give a simple phenomenological
describing of the base-stacking interaction strength as

\begin{equation}
K(f,\sigma)=-(0.1\sigma-0.01)e^{\alpha f+\gamma}-0.6(\alpha
f+\gamma)-2~, \label{k}
\end{equation}
where $\sigma$ denotes the supercoiling degree defined as
$$\sigma=\frac{L\kappa-L\kappa_0}{L\kappa_0}~,\label{x}$$
here $L\kappa$ is the linking number of the two strands whose
value is the sum of the twist and the writhe, $L\kappa_0$ is the
linking number of two strains in natural state.

The average unit bending persistence length  can be calculated by
the formula

\begin{equation}
\begin{array}{l}
 l=\displaystyle\frac{b}{2}\langle s+\tau\rangle+\epsilon~,\\[0.5cm]
\langle
s+\tau\rangle=\displaystyle\frac{\displaystyle\sum_{s_i,\tau_i=\pm
1}(s_j+\tau_j) e^{-\beta E_i}}{z}~, \label{m1}
\end{array}
\end{equation}
where $b$ is the unit physical bending persistence length (takes the
value of $b=0.34$nm per base pair), and $z$ denotes the partition
function of DNA molecule

\begin{equation}
z=\sum_{s_i,\tau_i=\pm1} e^{-\beta E_i}~~~~~~\beta=1/k_B T~.
\end{equation}

 \begin{figure}[h]
 \begin{center}
 \includegraphics[scale=0.8]{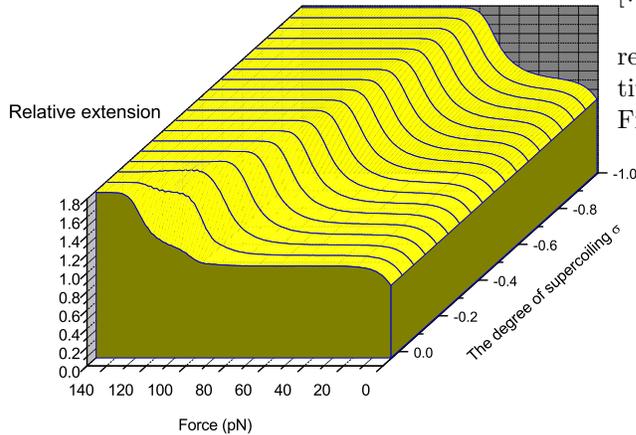}
 \caption[]{Relative extension of dsDNA molecule $vs$. stretching
 and twisting. The surface-graphics is obtained for
 $\alpha=0.04348$ and $\gamma=0.43478$. Other parameters
 are selected as: $L_{0}\approx15.1\mu$m (from Ref.\cite{st}),
 $\epsilon=0.04$nm, $N=71059$ and $b=0.34$nm as physical length per base pair.  }
 \end{center}
 \end{figure}
The total relative extension of dsDNA molecule under stretching
and twisting can be obtained

\begin{equation}
\frac{Nl}{L_0}=\frac{N}{L_0}\left(\frac{b}{2}\langle s+\tau
\rangle + \epsilon\right). \label{m2}
\end{equation}

In Fig. 3, we present a plot of the relative extension of dsDNA
molecule $vs$. stretching and twisting.

The surfacegraphics demonstrates clearly that the Ashkin-Teller
like model describes well the elastic response of dsDNA molecule
under both stretching and twisting simultaneously with  same
parameters for the B and S state when the DNA molecule stretched
over its contour length.

Fig. 3 exhibits three distinct  DNA states (corresponding to
different elastic property) and transitions among them induced by
external stretching. Furthermore, it reveals quantitatively that the
transition between B-form to S-state is influenced greatly by
external twisting. With the supercoiling $\sigma$ increasing from
negatively to positively supercoiled, the transition from the B form
to the S-state needs larger external force and the cooperative
transition becomes more sharpened. This is in agreement
quantitatively with experiments.

\begin{figure}[h]
\begin{center}
\includegraphics[scale=0.8]{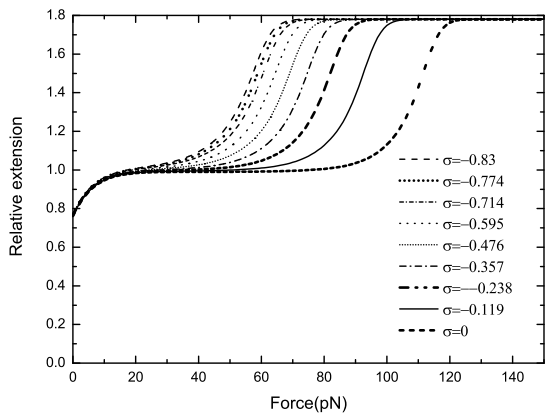}
\caption[]{Relative extension of the dsDNA molecule under stretching
with fixed twisting. The theoretic curves are obtained with the same
parameters in Fig.3. }
\end{center}
\end{figure}

To further investigate the relative extension of the DNA molecule
under stretching with fixed twisting and to compare with the
experiment data\cite{st}, we present the force-extension curves with
fixed supercoiling $\sigma$ in Fig. 4. It is in good agreement with
experimental observations \cite{st}.

The model gives also the asymmetric behavior of the relative
extension of the dsDNA molecule between negatively supercoiled and
positively supercoiled region (see Fig. 5).
\begin{figure}[h]
\begin{center}
\includegraphics[scale=0.8]{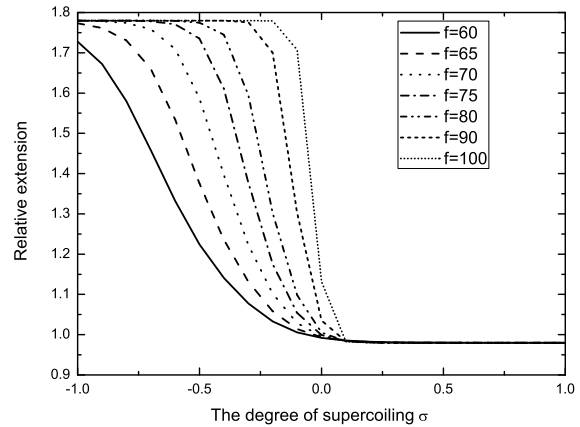}
\caption[]{Relative extension of the dsDNA molecule under twisting
with fixed force. The theoretic curves are obtained with the same
parameters in Fig. 3.}
\end{center}
\end{figure}
At the region of negatively supercoiled, the elastic response of
dsDNA behaves differently for different twisting  while keeping
force a constant. However, when $\sigma$ rises up to about $0.1$,
all curves get-together and have no difference in the course of
$\sigma$ increasing continuously. These results indicate that
twisting have much more influence on negatively supercoiled dsDNA
than on positively supercoiled one . The realistic significance of
the exotic behavior for dsDNA molecule is that, in the living, cell
dsDNA is usually negatively supercoiled ($\sigma\approx -0.06$).
Experiments\cite{strick} really showed asymmetric behavior of
elastic response of dsDNA molecule when external force is under $8$
 pN and the stretched dsDNA molecule is smaller than its contour
length. Our model reveals that the same property still exists after
dsDNA molecule is stretched over its contour length.  We wish more
elaborate experiments should check the validity of the prediction.

\section{III. Folding angle calculation and  reducing description of ssDNA }
\begin{figure}[h]
\begin{center}
\includegraphics[scale=0.8]{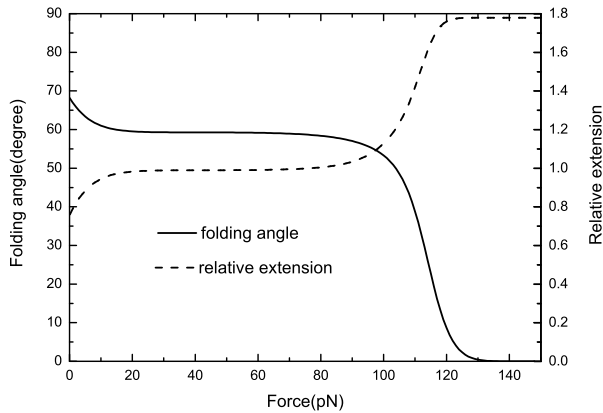}
\caption[]{Folding angle $vs$. relative extension under external
force(in the case of $\sigma=0$, $i.e$., torsion free). While the
dsDNA molecule is stretched to its contour length, the folding
angle has value $\varphi=59.2^{o}$, which is in good agreement
with the realistic structure of dsDNA (with
$\varphi\simeq62.0^{o}$).}
\end{center}
\end{figure}
Refs. \cite{zhou1,zhou2} postulated that dsDNA structure should be
extremely important to its elastic properties and introduced a
structural parameter: the folding angle $\varphi$, which describes
the angle between one of dsDNA backbones and central axis. In the
Ashkin-Teller like model, value of folding angle can also be
obtained

\begin{equation}
\langle\cos\varphi\rangle=\frac{b\langle s
\rangle}{b}=\frac{b\langle \tau \rangle}{b}.
\end{equation}
Fig. 6 demonstrates that the elastic behaviors of dsDNA molecule
under external force are really closely related to its structure and
the postulation \cite{zhou1,zhou2} is verified.

As shown above, the Ashkin-Teller like model describes well the
elastic behavior of dsDNA molecule. The base-stacking interaction
between the double stranded structure is represented by the strength
of ``four site'' interaction $K(f,\sigma)$ uniformly. We know that
this base-stacking interaction is the main factor maintaining the
stable double-helix structure and resulting in configurational
transformation. Without this interaction the model reduced to two
decoupled chains. The fact provides  possibility of describing
elastic response of dsDNS and ssDNA in a unified frame.

\begin{figure}[h]
\begin{center}
\includegraphics[scale=0.8]{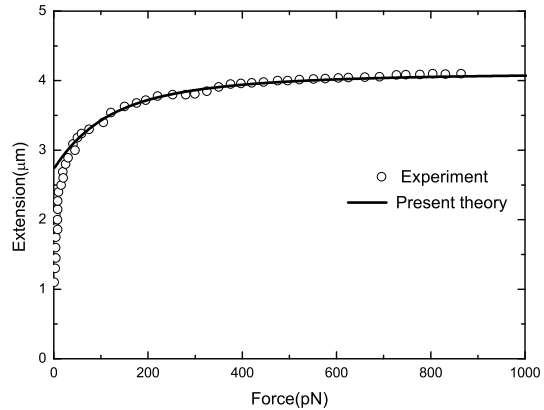}
\caption[]{Relative extension of ssDNA   under stretching. The
theoretic curves are obtained with  $\alpha=0.00086$ and
$\gamma=0.15437$. Other parameters are selected as: $L_{0}=3.9 \mu$m
(from Ref.\cite{rief}), $\epsilon=-0.35$nm, $N=3250$ and $b=0.6$nm
as its physical segment length. Experiment data are from
Ref.\cite{rief}.}
\end{center}
\end{figure}
Fig. 7 really shows a good agreement with ssDNA experiment data
\cite{rief}.  Other models \cite{nel} dealt with the elastic
response of dsDNA and ssDNA separately and have severe deviation in
describing of extension for ssDNA  when external force is larger
than $400pN$.

\section{IV. Conclusions}

The proposed Ashkin-Teller like model gives a quantitative
description for elastic response of dsDNA under both stretching and
twisting in a uniform way. It is in good agreement with experiment
data.  We obtained the phase diagram of dsDNA, and in particular,
the transition from B form to the S state induced by stretching and
twisting.  The close relation of dsDNA molecule structure with
elastic response has been shown clearly.

{\bf Acknowledgments:}  We thank Yanheng Li and Yue Yu for valuable
discussions. The work was supported in part by NSFC under Grants
10575106 and 10375072.


\end{document}